\documentstyle[aps,psfig]{revtex}
\draft
\tightenlines
\begin{document}
\title{A Knob for Changing Light Propagation from Subluminal to Superluminal}
\author{G.S.Agarwal$^{1}$, Tarak Nath Dey$^{1}$, Sunish Menon$^{1,2}$}
\address{$^{1}$Physical Research Laboratory, Navrangpura, Ahmedabad-380 009, India
\\ $^{2}$Intense Laser Physics Theory Unit and Department of Physics, 
Illinois State University, Normal, IL 61790-4560}
\date{\today}
\maketitle
\begin{abstract}
We show how the application of a coupling field connecting the two lower
metastable states of a $\Lambda$-system can produce a variety of new results on
the propagation of a weak electromagnetic pulse. In principle the light
propagation can be changed from subluminal to superluminal. The  negative group
index results from the regions of anomalous dispersion and gain in susceptibility.
\end{abstract}

\newpage
A series of experiments have demonstrated both subluminal \cite{Kas,Sch,Hau,Kash,Bud}
and superluminal \cite{Wang,Chu,A1} propagation of light in a dispersive 
medium. The key to these successful demonstration lies in one's ability to control
optical properties of a medium by a laser field. Harris et. al. \cite{Harr} 
suggested how the electromagnetically induced transparency (EIT) \cite{Harris} can be 
used to obtain group velocities $v_g$ \cite{Brill} much smaller than the velocity 
of light in vacuum. Early experiments \cite{Kas,Sch} produced values of group index
$n_g=c/v_g$ in the range $10^2 - 10^3$. Hau et. al. \cite{Hau} could reduce the 
group velocity to 17 meter/sec in a Bose condensate. This was followed by an 
experiment in Rb vapor demonstrating reduction of group velocity to $90$ 
meter/sec \cite{Kash} and to 8 meter/sec \cite{Bud}. These experiments were based 
on the fact that EIT not only makes absorption zero at the line center but also 
leads to a dispersion profile \cite{Harris,Xiao} with a sharp derivative near the 
line center of the absorption line. In a different development Wang et. al. \cite
{Wang} demonstrated superluminal propagation following the work of Chiao and 
coworkers [13-16; see also 17]. Wang et. al. used basically stimulated Raman effect
with the pumping beam replaced by a bichromatic field. This produces two regions 
of Raman gain with a region in between which has the right anomalous dispersion but
with a negligible gain \cite{Bow,Sunish}. In this communication we propose a 
scheme where by changing a knob - an additional coupling field, one can switch the propagation
of light from subluminal to superluminal. We basically use the $\Lambda$-system 
driven by a coherent control field, which has been extensively discussed in 
connection with subluminal propagation \cite{Kas,Sch,Hau,Kash,Bud}. We apply, in 
{\it addition}, a field (referred to as LL coupling field) on the lower levels of 
the $\Lambda$-system and demonstrate how the application of the lower
level coupling field 
can produce regions in optical response with appropriate dispersion profile. The
dispersion can change {\it from normal to anomalous} depending on the intensity 
of the LL coupling field \cite{Aldo,Bort,A2}. In addition,  under suitable conditions the
amplification of the light remains negligibly small.

We consider the scheme shown in the Fig. 1(a). We consider propagation of light
pulse whose central frequency $\omega_1$ is close to the frequency of the atomic
transition $|1 \rangle \leftrightarrow |3 \rangle$ . We apply a control field on the optical
transition $|1 \rangle \leftrightarrow |2 \rangle$. The transition 
$|2 \rangle \leftrightarrow |3 \rangle$
is generally electric dipole forbidden transition. The states $|2 \rangle$ and $|3 \rangle$
are metastable states. We apply a field of frequency $\omega_3$ on the
transition $|2 \rangle \leftrightarrow |3 \rangle$. The nature of this field will
depend on the level structure. It could be a microwave field, say, in case of Na
or an infrared field in case of $^{208}$Pb. Moreover, it could be a dc field
if one is considering transparency with Zeeman sublevels\cite{Bud}.
 Let $2G=2\vec{d}_{12}.\vec{E}_c/\hbar
~$ and $~2\Omega$ be the  Rabi frequencies of the control
field $\vec{E}_c$ and the LL coupling field, respectively. The state $|1
\rangle$ decays to the states
$|3 \rangle$ and $|2 \rangle$ at the rates $2\gamma_1$ and $2\gamma_2$. For simplicity we
ignore all collisional effects though these could be easily included. What is
relevant for further consideration is the group velocity $v_g$ for the pulse
applied on the transition $|1 \rangle \leftrightarrow |3 \rangle$. The $v_g$ is related to the
susceptibility $\chi_{_{13}}(\omega_1)$ for transition $|1 \rangle \leftrightarrow |3 \rangle$
\begin{equation}
v_g=\frac{c}{1+2\pi{\chi^{'}_{_{13}}}(\omega_1)+2\pi\omega_1\frac{\partial
{\chi^{'}_{_{13}}}(\omega_1)}{\partial \omega_1}}~~,
\end{equation}
where ${\chi^{'}_{_{13}}}(\omega_1)$ is the real part of $\chi_{_{13}}(\omega_1)$. We assume
that we are working under condition such that
Im$[\chi_{_{13}}(\omega_1)]=\chi''_{_{13}}(\omega_1)\approx 0$. The susceptibility
$\chi_{_{13}}(\omega_1)$ will depend strongly on the intensities and the frequencies of
the control laser and the LL coupling field. We concentrate on the group velocity though 
actual pulse profiles could be easily simulated \cite{Gar}. This susceptibility
$\chi_{_{13}}(\omega_1)$ is
obtained by solving the density matrix equations for the $\Lambda$-system of
Fig. 1(a), i.e., by calculating the density matrix element $\rho_{13}$ to first order in
the applied  optical field on the transition $|1 \rangle \leftrightarrow |3 \rangle$ but to all orders
in the control field and the LL coupling field. By making a unitary transformation from
the density matrix $\rho$ to $\sigma$ via 
\begin{equation}
\rho_{12}=\sigma_{12}e^{-i\omega_2
t}~~~;~~~\rho_{13}=\sigma_{13}e^{-i(\omega_2+\omega_3)t}~~~;~~~\rho_{23}=\sigma
_{23}e^{-i\omega_3 t}~~,
\end{equation}
we have the relevant density matrix equations
\begin{eqnarray}
\dot{\sigma}_{11}&=&i G \sigma_{21}+i g e^{-i\Delta_4
t}\sigma_{31}-i G^{*} \sigma_{12}-i g^{*} e^{i\Delta_4
t}\sigma_{13}-2(\gamma_1+\gamma_2)\sigma_{11}~,\nonumber\\
\dot{\sigma}_{22}&=& i G^* \sigma_{12} + i \Omega \sigma_{32} - i G \sigma_{21}
- i \Omega^* \sigma_{23} + 2\gamma_2\sigma_{11}~,\nonumber\\
\dot{\sigma}_{12}&=&-[\gamma_1+\gamma_2+\Gamma_{12}-i\Delta_2]\sigma_{12}+i G
\sigma_{22}+ig e^{-i\Delta_4 t}\sigma_{32}-iG\sigma_{11}-i \Omega^*
\sigma_{13}~,\\
\dot{\sigma}_{13}&=&-[\gamma_1+\gamma_2+\Gamma_{13}-i(\Delta_2+\Delta_3)]\sigma_{13}
+iG\sigma_{23}+ige^{-i\Delta_4 t}\sigma_{33}-ige^{-i\Delta_4
t}\sigma_{11}-i\Omega\sigma_{12}~,\nonumber\\
\dot{\sigma}_{23}&=&-(\Gamma_{23}-i\Delta_3)\sigma_{23}+iG^*\sigma_{13}+i\Omega
\sigma_{33}-ige^{-i\Delta_4 t}\sigma_{23}-i\Omega\sigma_{22}~,\nonumber
\end{eqnarray}
where $\Gamma$'s give collisional dephasings, the detunings $\Delta_1,~ \Delta_2,
~\Delta_3, ~\Delta_4$ and the coupling 
constant$~g$ are defined by 
\begin{equation}
\Delta_1=\omega_1-\omega_{13}~;~\Delta_2=\omega_2-\omega_{12}~;~\Delta_3
=\omega_3-\omega_{23}~;~\Delta_4=\Delta_1-\Delta_2-\Delta_3~;~g=\frac{\vec{d}_{13}.\vec{E}_p}{\hbar}~~.
\end{equation}
The susceptibility $\chi_{_{13}}$ can be obtained by considering the steady state
solution of (3) to first order in the field on the transition $|1 \rangle
\leftrightarrow |3 \rangle$. For this purpose we write
\begin{equation}
\sigma=\sigma^0~+~\frac{g}{\gamma}~e^{-i\Delta_4
t}~\sigma^+~+~\frac{g^*}{\gamma}~e^{i\Delta_4 t}~\sigma^-+......
\end{equation}
The $13-$ element of $\sigma^+$ will yield the susceptibility at the frequency
$\omega_1$ as can be seen by combining Eqs. (2) and (5)
\begin{equation}
\chi_{_{13}}(\omega_1)=\frac{N|d_{13}|^2}{\hbar \gamma}\sigma_{13}^+~~,
\end{equation}
where $N$ is the density of atoms. In the above equations we have, for simplicity,
set $\gamma_1=\gamma_2=\gamma$. The group velocity can be obtained by
substituting (6) in (1). In presence of the LL coupling field it is difficult to
obtain algebraically simple expressions for $\chi_{_{13}}$. However Eqs. (3) can be
solved numerically. Doppler broadening can be accounted for by using
 $\omega_1 \rightarrow \omega_1-kv~;~\omega_2 \rightarrow
\omega_2-kv~;$ and by averaging over the Maxwellian distribution for velocities.
The velocity dependence of $\omega_3$ is insignificant and hence dropped. The
parameter $\delta=\left(\frac{k_B T \omega_1^2}{Mc^2}\right)^{1/2}$ is a measure of Doppler
width in the  Maxwellian distribution $p(x) \propto \exp(-\frac{x^2}{2\delta^2})~;~x=kv$. We show a number of
numerical results in Figs. 1 and 2. We notice from the Fig. 1(c) how the group index
 $n_g$ defined via $v_g=c/n_g$, changes from
large positive values to large negative values and back to positive values as
the intensity of the LL coupling field is increased. Thus the LL coupling field
is like a knob which can be used to change light propagation from subluminal to
superluminal. We also present the behavior of the corresponding susceptibility for parameters
corresponding to superluminal propagation in the Fig. 1(b). We see that at $\Delta_1=0$, the 
real part of $\chi_{_{13}}$
exhibits anomalous dispersion whereas the imaginary part of $\chi_{_{13}}$ is fairly
flat and negative and is exactly zero at $\Delta_1=0$.
The anomalous dispersion along with negative flat region in the imaginary part of
$\chi_{_{13}}$ is especially attractive for superluminal propagation \cite{tar}.
In Fig. 1(d) we show
the behavior of a pulse  ${\cal E}(t-L/c)=\frac{{\cal E}_0}{2\pi}\exp{\left[-(t-L/c)^2/\tau^2\right]}$,
${\cal E}(\omega) =\frac{{\cal E}_0}{\sqrt{\pi\Gamma^2}}
\exp{\left[-(\omega-\omega_0)^2/\Gamma^2\right]}$,
$\Gamma \tau = 2$, at  the output of a medium under condition that group
index is negative. The Fig. 1(d) shows that there is no distortion of the pulse. 
For comparison we also show the pulse at the output in the absence
of the medium. The advancement of the pulse due to medium is seen. The
difference in the peak positions is in agreement with negative value of the
group index. In Fig. 2 we show the results for the group index with and without
Doppler averaging. It is known from the work of Kash et. al.\cite{Kash} that Doppler broadening
was insignificant in the behavior of the pulse propagation through a
$\Lambda$-system in presence of a control laser. However the situation changes
with the application of the LL coupling field at $|2 \rangle \leftrightarrow |3 
\rangle $ 
transition (with a wavelength $\sim 1.3 \mu m$) particularly in the region where
group index is negative. In Fig. 2 we also show the results for propagation in a
much heavier $^{208}$Pb vapor . This was the system used earlier by Kasapi et.
al. \cite{Kas} to demonstrate subluminal propagation. The application of the LL
coupling field
can lead to the superluminal propagation. The results in this case are not
sensitive to Doppler broadening because the Doppler width parameter $\delta~(\cong 
25 \gamma)$ is much smaller than the Rabi frequency $G~(=297\gamma)$ of the
pump. We note that the production of superluminal
propagation depends very much on the nature of the atomic transitions
in the system under study and the choice of a large number of parameters such as
the powers of the control and coupling fields. From our numerical results it is clear 
that
we need a large coupling between $| 2 \rangle$ and $| 3 \rangle$. For a magnetic
dipole transition between the states $| 2 \rangle$ and $| 3 \rangle$ the
requirement of power of the LL coupling field is large and, in principle, this can
be met by using pulsed fields with a pulse width $\gtrsim \mu$ sec.
However if $| 2 \rangle$ and $| 3 \rangle $ are chosen to be Zeeman levels, then
the available dc magnetic field can be utilized to change propagation from subluminal to
superluminal. Note that for Rb, a Rabi frequency of $100\gamma$ implies a
magnetic field of the order of $99.3$ Gauss. Another
possibility would be to consider an effective interaction between $| 2 \rangle$
and $| 3 \rangle $ via Raman transition using two other laser fields. The choice
of the system is quite open and we have essentially shown the ``in principle''
possibility of light propagation from subluminal to superluminal. Thus in
conclusion we have demonstrated how the $\Lambda$ system can produce a variety
of new results if we apply an additional LL coupling field. In particular we have
demonstrated how the application of the LL coupling can produce regions of
anomalous dispersion with gain and how this results in superluminal propagation
of a weak pulse of light.\\[20pt]
One of us (GSA) thanks Ennio Arimondo and Steve Harris for interesting
suggestions and S. Menon acknowledges the NSF support from grant no. PHY-9970490.

\begin{figure}
\vspace* {-.9cm}
\centerline{\begin{tabular}{cc}
\psfig{figure=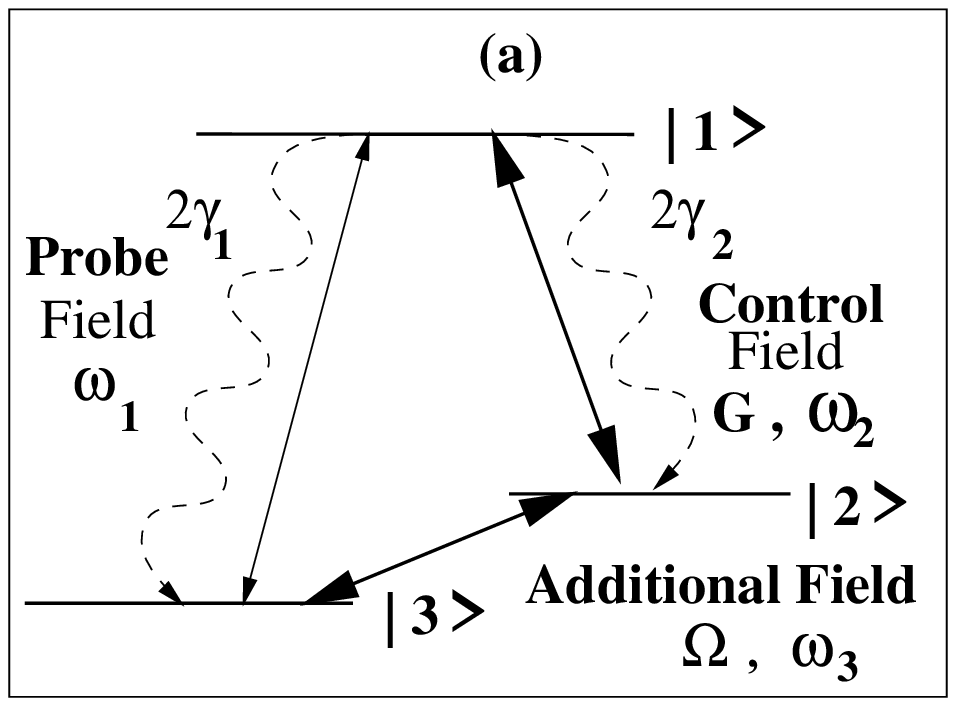,height=8.0cm,width=9.5cm}&
\psfig{figure=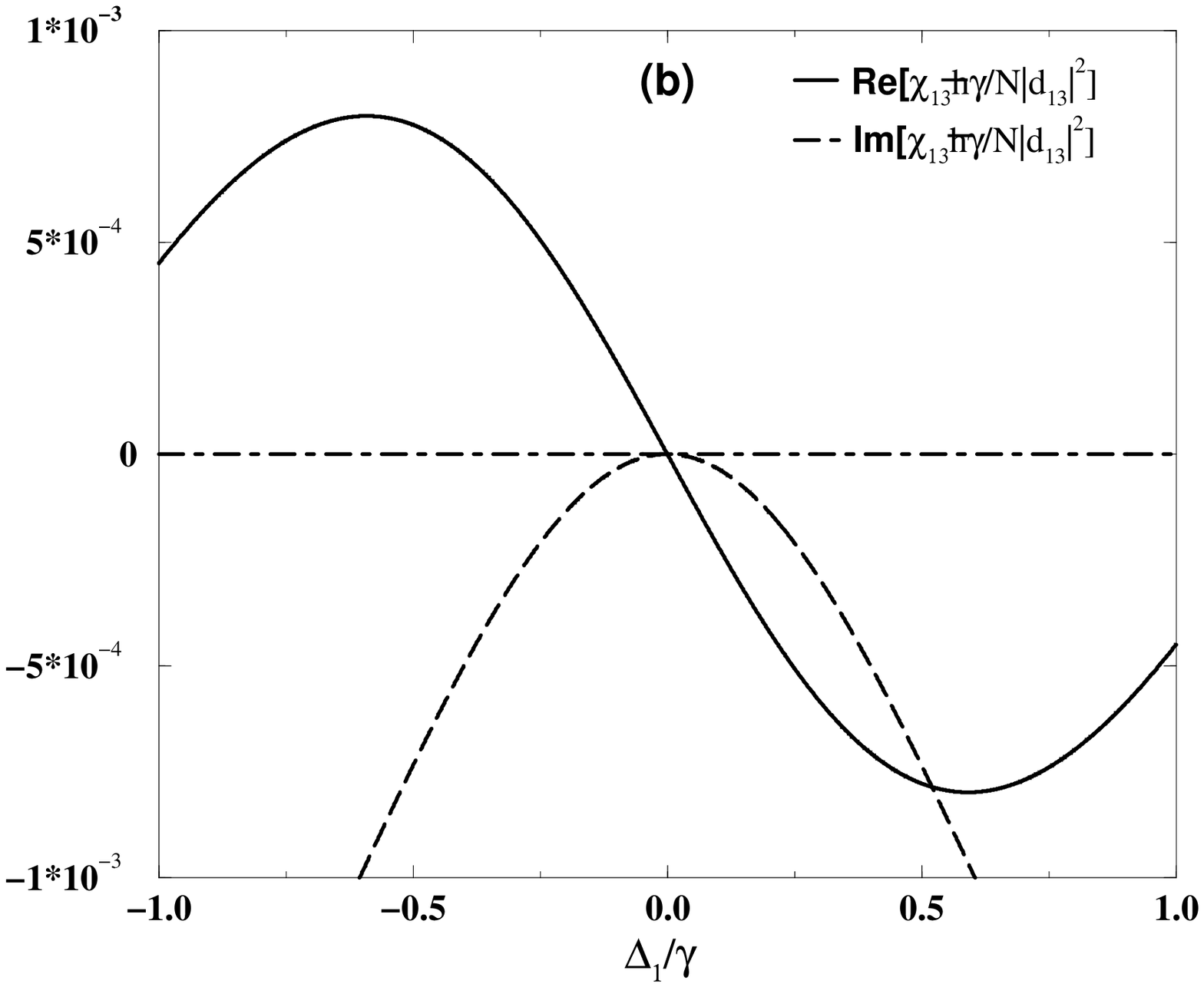,height=7.5cm,width=9.5cm}\\[1pt]
\psfig{figure=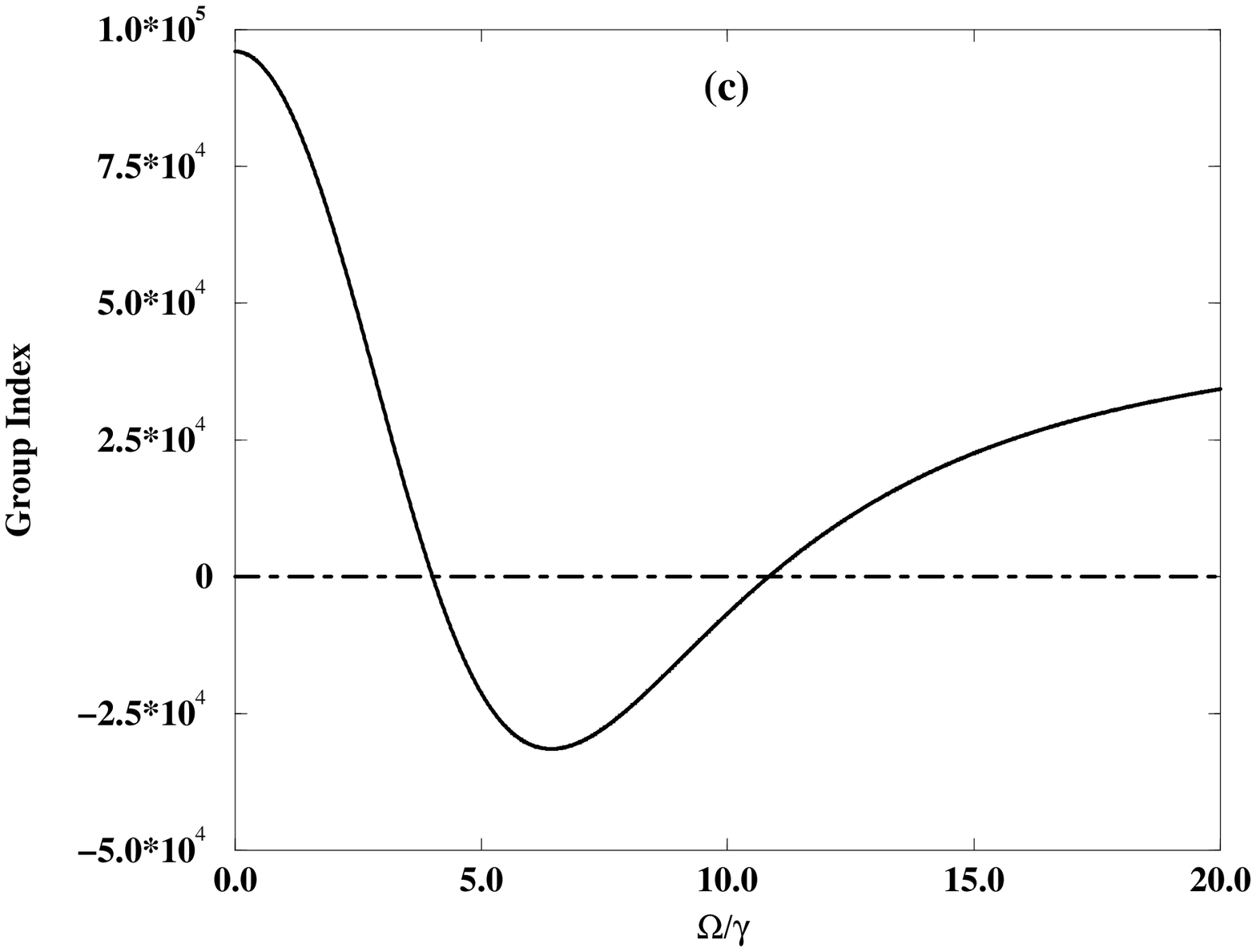,height=7.5cm,width=9.5cm}&
\psfig{figure=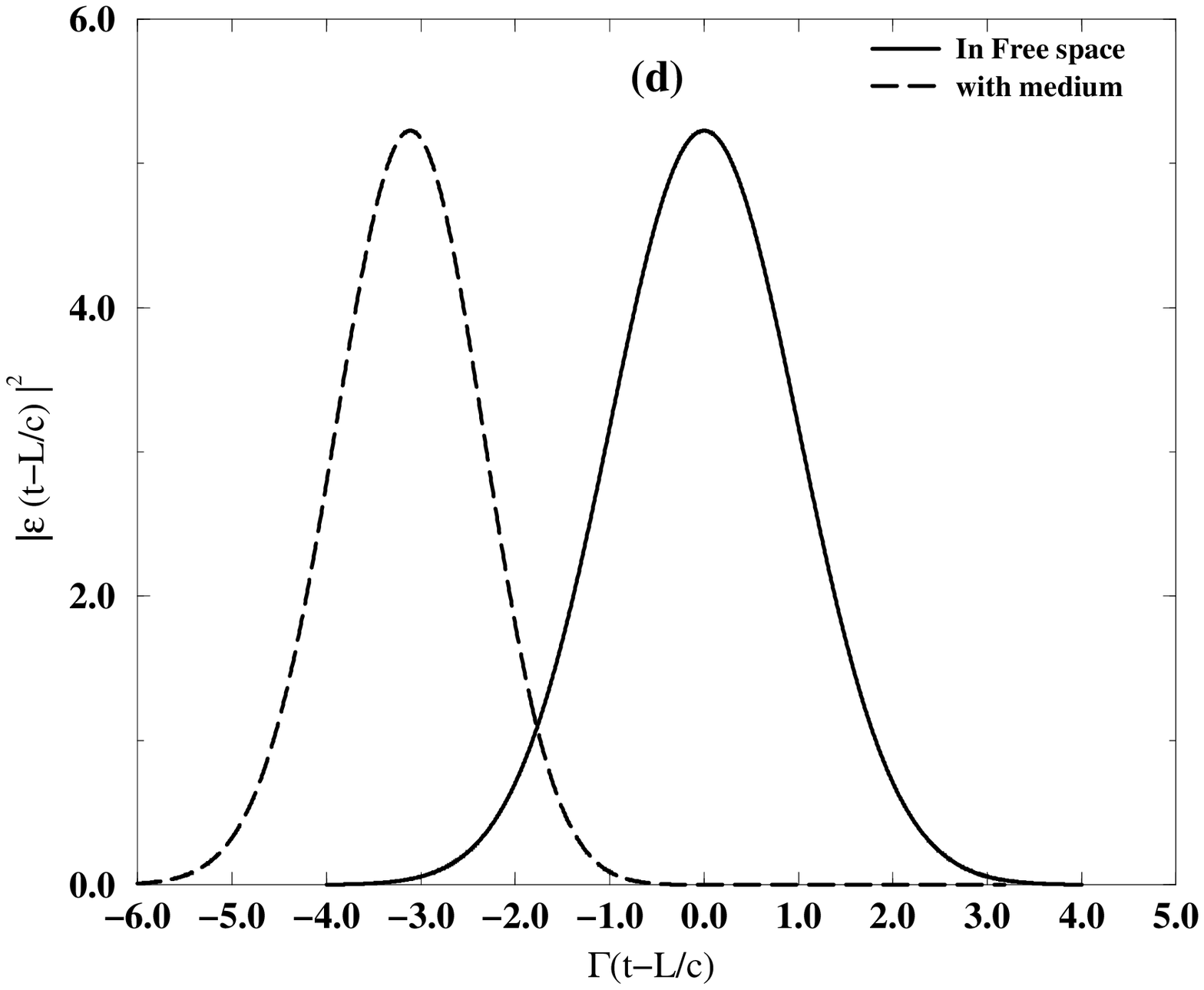,height=7.5cm,width=9.5cm}
\end{tabular}}
\caption{(a) Schematic diagram of three level $\Lambda$-system. (b) Real and imaginary parts of the susceptibility [Eqs. (6) ;
$\chi_{_{13}}\hbar\gamma/\textrm{N}|d_{13}|^2$] at a
probe frequency $\omega_1$ in the presence of control and LL coupling field $\Omega=5\gamma$. (c)
Variation of group index with the Rabi frequency of LL coupling field. The pulse
is taken to have a central frequency on resonance with the transition 
$|1 \rangle \leftrightarrow |3 \rangle$.
The solid curve of (d) shows light pulse propagating at speed $c$
through 6 cm of vacuum. The dotted curve shows same light pulse propagation
through the medium of length 6 cm with a time delay $-4.39\mu$sec 
in presence of a LL coupling field with Rabi frequency  $\Omega=5\gamma$. The pulse
width $\Gamma$ is $120$KHz. The common
parameters of the above three graphs for $^{87}$Rb vapor are chosen as
density
N$=2*10^{12}$ atoms/cc, G$=10\gamma$, 
$\Delta_2=\Delta_3=0$, $\Gamma_{12}=\Gamma_{13}=\Gamma_{23}=0$, $\gamma=3\pi*10^{6}$ rad/sec 
. }
\end{figure}
\begin{figure}
\centerline{\begin{tabular}{c}
\psfig{figure=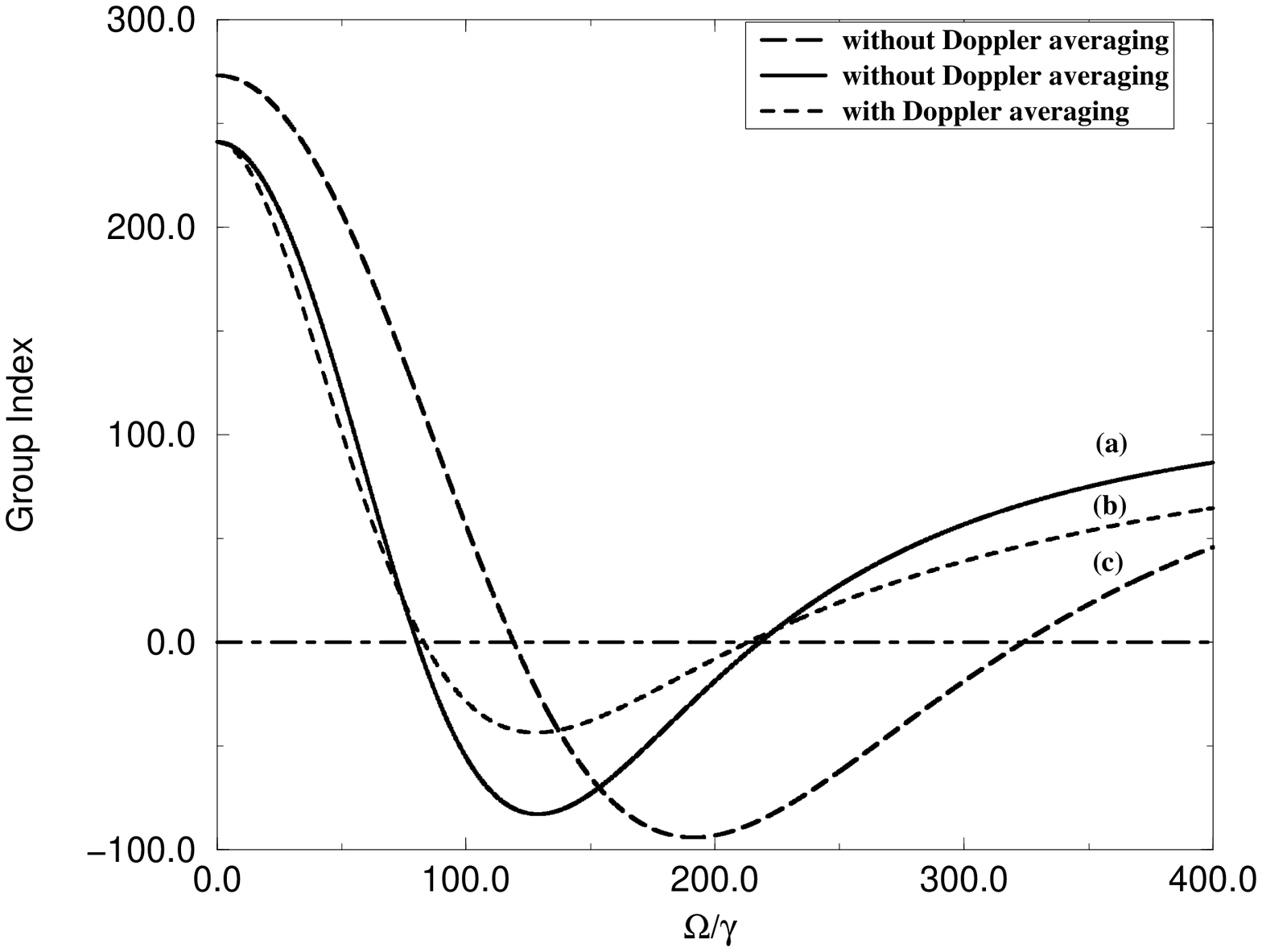}
\end{tabular}}
\caption{ Group index variation with the 
Rabi frequency of LL coupling field. The curves (a) and (b) are for propagation
in Rubidium vapor with density N$=2*10^{12}$ atoms/cc. Other parameters are chosen as
G$=200\gamma$, $\Delta_1=\Delta_2=\Delta_3=0$, $\Gamma_{12}=\Gamma_{13}=\Gamma_{23}=0
$, $\gamma =3*\pi*10^{6}$ rad/sec. For the curve (b) the Doppler
width parameter $\delta$ is chosen as $1.33*10^{9}$ rad/sec. Curve (c) shows variation of
group index $n_g$ with the Rabi frequency of LL coupling field in $^{208}$Pb vapor
with density 
N$=2*10^{14}$ atoms/cc, G$=297\gamma$, $\Delta_1=\Delta_2=\Delta_3=0$,
$\Gamma_{12}=\Gamma_{13}=\Gamma_{23}=0$, $\gamma=4.75*10^{7}$ rad/sec.}
\end{figure}


\begin{thebibliography}{999}
\bibitem{A2} Budker et. al. noted in Ref. 5 how the propagation can be changed
from subluminal to superluminal by using, say, static magnetic fields of the order of
few $\mu G$.
\bibitem{tar} In the experiment of Wang et. al. similar regions of 
$\chi_{_{13}}$ were used to produce superluminal propagation.
\end{thebibliography}
\end{document}